\documentclass{llncs}

\usepackage[utf8]{inputenc} 
\usepackage[T1]{fontenc}    
\usepackage{lmodern}
\usepackage{hyperref}       
\usepackage{url}            
\usepackage{booktabs}       
\usepackage{amsfonts}       
\usepackage{nicefrac}       
\usepackage{microtype}      
\usepackage{amsmath}
\usepackage{graphicx}
\usepackage{mathrsfs}
\usepackage{algorithm}
\usepackage{algorithmic}
\usepackage{array}
\usepackage{amsmath}
\usepackage{graphicx}
\usepackage{amsfonts,color}
\usepackage{subcaption}
\usepackage{diagbox}
\usepackage{xcolor,todonotes,wrapfig,colortbl}

\usepackage{tikz}
\usepackage{pgfplots}
\usepackage{pgfkeys}
\usetikzlibrary{spy}
\usetikzlibrary[patterns] 
\usetikzlibrary{calc,decorations.pathreplacing} 
\usetikzlibrary{shadows.blur} 
\usetikzlibrary{shapes.symbols}

\DeclareMathOperator*{\argmin}{arg\,min}

\DeclareMathOperator*{\argmax}{arg\,max}

\title{Learning local regularization for variational image restoration}

\author{Jean Prost\inst{1}\and
Antoine Houdard\inst{1}\and
Andrés Almansa\inst{2}\and 
Nicolas Papadakis\inst{1}}
\institute{Univ.  Bordeaux, Bordeaux INP, CNRS, IMB, UMR 5251,F-33400 Talence, France
\email{jean.prost@math.u-bordeaux.fr}\\
\and
Université de Paris, MAP5, CNRS, F-75006 Paris, France\\
}

\begin{document}
\pagestyle{plain}

\maketitle

\begin{abstract}
In this work, we propose a framework to learn a local regularization model for solving general image restoration problems. This regularizer is defined with a fully convolutional neural network that sees the image through a receptive field corresponding to small image patches. The regularizer is then learned as a \emph{critic} between unpaired distributions of clean and degraded  patches using a Wasserstein generative adversarial networks based energy. This yields a regularization function that can be incorporated in any image restoration problem. The efficiency of the framework is finally shown on denoising and deblurring applications.

\end{abstract}
\section{Introduction} 
\label{sec:intro}

\paragraph{Inverse problems and convex regularization.}

Many image restoration tasks require to solve an inverse problem. This can be addressed with a variational formulation involving a data-fidelity term and a regularization term encouraging the solution to satisfy given properties or to belong to a space of possible solutions. Some of the most famous regularization terms used for image restoration are convex non-smooth terms like the total variation~\cite{rudin1992nonlinear}, or $\ell^1$ minimization of transform-domain coefficients such as Wavelet frames \cite{Donoho1994,Coifman1995} or local Fourier or DCT representations \cite{Yu2011}. However theses strategies tend to produce over-smoothed results, since they represent only a rough approximation of natural image statistics and geometry.

\paragraph{CNN-based non-convex regularization.}
Later-on more accurate natural image priors emerged in the form of non-convex regularization terms, such as patch-based Gaussian mixture models (to be discussed below) or convolutional neural networks (CNN). Most common CNN-based regularizers are, however, trained in a way that the prior or regularizer itself is only partially and implicitly known via its gradient \cite{Bigdeli2017,Romano2016red,Reehorst2018a} or proximal operator \cite{venkatakrishnan2013plug,meinhardt2017learning,Zhang2017,kamilov2017plug,ryu2019plug}. Such implicit CNN regularizers, and the associated optimization algorithms, lack convergence guarantees or do so under overly restrictive conditions on the regularizer, the regularization parameter or the kind of inverse problems they can solve \cite{Reehorst2018a,ryu2019plug}.

To overcome these limitations a new breed of explicit CNN-based regularizers have been proposed, either in the form of the push-forward measure of a generative model~\cite{bora2017compressed}, a variational autoencoder~\cite{Gonzalez2019}, or more directly as a discriminator network~\cite{lunz2019}. All these approaches are nevertheless limited to a particular class of image and do not generalize to images of arbitrary size.

\paragraph{Patch-based non-convex regularization.}

Learning prior information has also been widely studied from the patch point-of-view. The main idea is to learn the prior knowledge from patches, that are local sub-images of small size, instead of learning a prior from whole images. This allows to avoid the high-dimensional issues faced when working with full-size image distributions. These approaches rely on parametric models of the patch distribution such as Gaussian mixture models~\cite{EPLL,HDMI,Teodoro2018scene}. However such simple models can not accurately represent the complexity of the patch space.

In this work, we introduce an explicit non-convex regularization function encoded with a fully convolutional neural network that acts as a local regularizer. This prior knowledge on the patch distribution can be applied to a whole image without size limitation. We propose (i) to learn the convolutional regularizer as a discriminator between patches using the Wasserstein GAN framework~\cite{WGAN} as in~\cite{lunz2019}, and (ii) to integrate this regularizer in patch-based models such as~\cite{EPLL}.

\subsection{Setup of the problem}
The main goal of this paper is to perform image restoration by solving an inverse problem. That is, finding the underlying true image $x^\star$ from its perturbed observation $y$ that we consider here to be of the form 
\begin{equation}
\label{eq:degrad}
y = Ax^\star+\epsilon,
\end{equation}
where $\epsilon\sim \mathcal{N}(0,\sigma^2)$ is a Gaussian white noise and $A$ is a degradation operator that can typically be the identity (pure denoising), a mask (inpainting) or a blurring kernel (deconvolution). These inverse problems can be addressed with a variational formulation involving a regularization term. This amounts to find an estimate $\hat{x}$ of $x^\star$ of the form
\begin{equation}
\label{eq:var}
\hat{x} \in \argmin_x \frac{1}{2\sigma^2}\|Ax-y\|^2 + \lambda R(x),
\end{equation}
where  $\|Ax-y\|^2$ is the data-fidelity term ensuring that the recovered image $\hat{x}$ is close enough to the degraded observation $y$, $R(x)$ is the regularization term and $\lambda\geq 0$ monitors the influence of both terms. In the case where $R(x) = -\log(P_X(x)) + C$ is derived from a prior probability distribution $P_X$ modeling the data $x$, then $\hat{x}$ from \eqref{eq:var} corresponds to the \emph{maximum a posteriori} estimator.

The choice of the regularization function $R$ has a strong impact on the final result. We propose to learn $R$ through a local regularization functional $r$ acting on patches. Denoting as $\Omega_x=\{x_1,\cdots, x_n\}$ the set of all patches of size $p\times p$ from an image $x$, this function takes as input an image patch $x_i$ and outputs a score $r(x_i)$ that indicates how likely the patch is to be a clean one. As  in~\cite{EPLL}, we define the global regularization functional as the average value of the local scores on the set of all patches of image $x$:

\begin{equation}
\label{eq:single_scale}
R(x) = \frac{1}{|\Omega_x|}\sum_{x_i \in \Omega_x }r(x_i).
\end{equation}


Working with patches yields three main advantages. It first makes the learning phase simpler, as a patch model contains far less parameters than a full image model. Next, the number of images required for training is reduced, as a single image already provides several thousands of patches. Finally, contrary to~\cite{lunz2019}, the regularization function can be applied on images of any size.

In practice, we  consider $r$ as a CNN with perceptual size equal to the patch size $p\times p$ and taking values in $\mathbf{R}$. 
This representation is more general than Gaussian mixture models, and allows to encode complex distributions.

\subsection{Contributions and outline.}
We propose an image restoration method that relies on a regularization function learned on patches and applied to any image size. It gathers the advantages of previous CNN methods while avoiding the constraints of implicit plug \& play priors (convergence guarantee) and of GAN or VAE priors (image size).

In addition, the regularization function is learned in an \emph{unsupervised} manner, in the sense that it only relies on  patch distributions of clean and degraded data and {\bf it does not require paired data}. We can therefore deal with an unknown degradation model if a noisy dataset is available.

The organization of the paper is as follows. 
In Section \ref{sec:patch_prior}, we propose an unsupervised framework for the learning of a compact convolutional neural network modeling the local patch regularity prior.
We namely obtain the local regularization functional $r$ as a critic trained to distinguish noisy patches from clean ones using the framework of Wasserstein generative adversarial models~\cite{WGAN}. 
In section \ref{sec:practical}, we provide implementation details to make the work fully reproducible.
We show in section \ref{sec:generalization} that the local functional $r$ generalizes well to arbitrary levels of noise, i.e. noise level unseen during training.
In Section \ref{sec:experiment}, we  demonstrate that the proposed framework is efficient for image denoising  and deblurring.

\section{Local regularization for image inverse problem}\label{sec:patch_prior}
In this section we define our local image regularizer $r_{\theta}$ as a convolutional neural network and we describe how we use and train it.

Patch-based methods have shown to be efficient tools for solving inverse problems in imaging~\cite{EPLL}. Hence we aim at defining a regularization function $r_\theta$ depending on parameters $\theta\in\Theta$ that encode prior knowledge at a patch level. In the patch-based literature, such regularizers rely on statistical modeling of the distribution of clean patches and the model parameters are usually inferred with a maximum likelihood estimation~\cite{HDMI}. This leads to two main limitations. First, it requires to have access to the probability density function of the prior distribution and consequently it does not  properly represent the intrinsic low dimensional manifold of clean patches. Second,  maximizing the likelihood of a complex model leads to non-convex problems that are difficult to solve in practice.

In order to tackle these issues, we propose to take advantage of having two data sets of clean and degraded patches --not necessarily paired-- and consider $r_\theta$ as a \emph{critic} that tells us if a patch is more likely to be clean or degraded.

We first detail in section~\ref{sec:using}  how the local regularization function is integrated as a global regularizer on images in order to solve the variational problem~\eqref{eq:var}. 
In  section~\ref{sec:learning}, we present the framework to learn the regularizer as a \emph{critic}  between two unpaired dataset of clean and degraded images.

\subsection{Convolutional regularizers for variational problems\label{sec:using}}

We define, for the variational problem~\eqref{eq:var}, a regularization term $R$ that takes into account local prior knowledge of the images. To do so, we propose to consider a class of functions $r_\theta$ defined with a fully convolutional neural network with parameter $\theta \in\Theta$. We enforce the perceptual size of this network to be the patch size $p\times p$. That is, the successive convolutions operate on a window no larger than $p\times p$ pixels. Using this  architecture permits to compute the global regularizer $R$ from~\eqref{eq:single_scale} by directly applying $r_\theta$ to the full image $x$ and average the outputs. 
Once learned the local regularizer $r^\star_\theta$, the variational problem to solve becomes 
\begin{equation}\label{eq:pb}
\min_x \frac{1}{2\sigma^2}\|Ax-y\|^2 + \frac{ \lambda}{|\Omega_x|}\sum_i r^\star_\theta(x_i).
\end{equation}
We propose to find a local minimizer of \eqref{eq:pb} by performing an explicit gradient descent method. Let $x^\ell$ the image at iteration $\ell$, a gradient step of step size $\eta$ writes
\begin{equation}
x^{\ell+1} = x^{\ell} - \frac{\eta}{\sigma^2}A^*(Ax^\ell-y) - \frac{\eta \lambda}{|\Omega_x|} \sum_i \nabla r^\star_\theta(x^\ell_i),
\end{equation}
where $A^*$ is the adjoint operator of $A$. Contrary to plug \& play methods that rely on implicit schemes \cite{venkatakrishnan2013plug}, this explicit scheme  converges for differentiable regularization functions and adequate time steps.

We now describe how the framework for learning the local regularization function.

\subsection{Adversarial Local Regularizer (ALR) \label{sec:learning}}



In order to train $r_\theta$ as a critic between patch distributions, we consider the discriminator framework introduced for generative adversarial networks~\cite{GAN}, without the generator network. Such approach nevertheless results in a critic $r_\theta$ approximating the hard clustering between clean and degraded patches. It therefore induces steep gradients $\nabla r_\theta$ that may lead to numerical instabilities during the minimization of problem~\eqref{eq:var}.

As a consequence, we rather rely on the Wasserstein GAN~\cite{WGAN} formulation that amounts to approximate the optimal transport cost between the distribution of clean patches $\mathbb{P}_c$, and a distribution of degraded patches $\mathbb{P}_n$. Relying on the dual formulation of the optimal transport~\cite{santambrogio2015ot}, an optimal critic $r^\star_\theta$ is seen as a Kantorovitch potential and shall satisfy  \begin{equation}
\label{kantodual}
 r^\star_\theta \in \argmax_{\varphi  \in \mathrm{Lip}_1} \mathbb{E}_{z\sim \mathbb{P}_n}\left[\varphi(z)\right] - \mathbb{E}_{z\sim \mathbb{P}_c}\left[\varphi(z)\right].
\end{equation} 
Under the assumption that the support of the clean patches distribution $\mathcal{M}$ is compact \cite{lunz2019}, the solution of equation \eqref{kantodual} corresponds to the distance function to the clean data manifold $\mathcal{M}$. Each iteration of the gradient descent on equation \eqref{eq:var} thus brings our noisy data closer to the clean data. 

In practice, imposing a neural network to be 1-Lipschitz is a difficult task and we therefore use the formulation proposed in~\cite{WGANGP} that encourages the gradient norm to be close to $1$. This amounts to maximize the following quantity
\begin{equation}
\label{wgancrit}
D(\theta)= \mathbb{E}_{z\sim \mathbb{P}_n}\left[r_{\theta}(z)\right] - \mathbb{E}_{z\sim \mathbb{P}_c}\left[r_{\theta}(z)\right] - \mu \mathbb{E}_{z\sim \mathbb{P}_i}[(||\nabla_z r_{\theta}(z)||_2 -1)^2]
\end{equation}
where $\mathbb{P}_i$ is the distribution of all lines connecting samples in $\mathbb{P}_n$ and $\mathbb{P}_c$. In other words, the last term of \eqref{wgancrit} is a gradient penalty that makes the function  1-Lipschitz on the convex hull of the union of the support of $\mathbb{P}_c$ and $\mathbb{P}_n$. By enforcing the gradient $\nabla r_{\theta}$ to be of norm close to $1$, vanishing gradient issues are also avoided when solving  problem \eqref{eq:var} with gradient descent approaches. 

\begin{figure}
\centering
\includegraphics[width=0.85\linewidth]{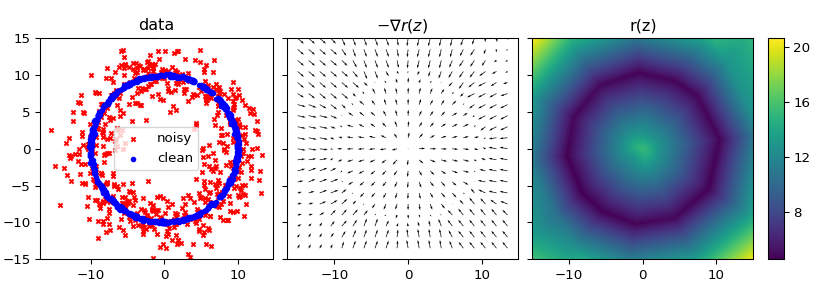}
\caption{Regularization functional $r(z)$ learned on a synthetic 2d denoising problem with clean data (blue dots) on the circle and noisy ones (red crosses). The gradient penalty ensures that the gradient $\nabla r$ is not flat close to the data manifold $\mathcal{M}$.}
\label{fig:2dcritic}
\end{figure}

We illustrate the properties of the regularization functional with a synthetic example in Figure~\ref{fig:2dcritic} containing random perturbations of clean data points located on a circle. The learned regularization function $r_\theta(z)$ therefore approximates the distance function to the circle. The gradient $\nabla r_\theta(z)$ thus indicates the direction to follow in order to transport $z$ towards the a clean point within the circle.

\section{Practical considerations for image restoration}
\label{sec:practical}
In this section, we provide implementation details to reproduce the proposed framework.  After presenting the architecture of the regularization network $r_\theta$, we  explain the training strategy and   describe how image restoration is performed.
\subsection{Network architecture}
The local regularization functional $r_{\theta}$ is designed as a 6 layers convolutional network. Each layer is made of $3 \times 3$ convolution operations followed by ReLU activations \cite{nair2010rectified}. This network has therefore a $15 \times 15$ receptive field. No padding is used. Hence, when a patch of the size of the network receptive field is fed to the network $r_{\theta}$, the output is a scalar.

\subsection{Training the regularization functional}
The proposed regularization network is trained with patches matching the size of the receptive field of the network. We create the dataset $\mathcal{D}_c$ of clean patches by extracting all $15 \times 15$ patches from a 30000 image subset of the google landmarks dataset \cite{weyand2020google}. Similarly, we create the dataset $\mathcal{D}_n$ of noisy patches by extracting all $15 \times 15$ patches from another 30000 images subset of the landmarks dataset, to which we added an additive white Gaussian noise with standard deviation $\sigma_{train}$. Following \cite{lunz2019}, the local regularization network $r_{\theta}$ is trained to minimize the criterion \eqref{wgancrit} with Algorithm \ref{alg:lunz}. We use the Adam optimizer \cite{kingma2017adam} with hyperparameters $\beta_1=0.9$ and $\beta_2=0.999$, and an exponential learning rate decay, so that the learning rate $\alpha$ begins at a value of $10^{-3}$ for the first iteration, and ends up at $10^{-4}$ for the last iteration. We use a batch size of $m=32$ and train the network for $K=10^5$ iterations. The gradient-penalty parameter is set to $\mu = 5$.

\begin{figure}[ht!]
\centering
\begin{tabular}{c}
\includegraphics[scale=2.5]{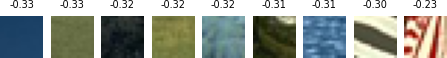}\\
\includegraphics[scale=2.5]{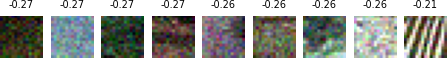}\\
\end{tabular}
\caption{Value of the local regularization functional $r_\theta$ trained with $\sigma_{train}=0.1$ on clean (top row) and noisy (bottom row) patches ($\sigma=0.1$).\vspace{-0.8cm}}
\label{fig:patches}
\end{figure}

Training samples of clean and noisy patches $z$,  with their final regularizer value $r_\theta(z)\in\mathbb{R}$, are shown in Figure \ref{fig:patches}. As can be observed from the functional values, there exists a slight ambiguity between texture patches ($r_\theta(z)=-0.23$ for the last patch of top row) and noisy homogeneous patches ($r_\theta(z)=-0.27$ for the first patches of bottom row).
We nevertheless show in Figure \ref{fig:hist} that the distributions of clean and noisy patches are globally well separated. 

\begin{algorithm}[t]\small
   \caption{Learning the local regularization $r_\theta$} 
   \label{alg:lunz}
\begin{algorithmic}
   \STATE {\bfseries Input:} Datasets $\mathcal{D}_c$ of clean patches and $\mathcal{D}_n$ of noisy patches; gradient penalty $\mu$; batch size $m$, number of iterations $K$
   \STATE {\bfseries Output:} regularization function $r_\theta$
   \FOR{$k=1$ {\bfseries to} $K$}
   \STATE Sample minibatches of $m$ clean patches $\{z^c_j\}_{j=1}^m$ from $\mathcal{D}_c$ and $m$ noisy patches $\{z^n_j\}_{j=1}^m$ from $\mathcal{D}_n$ and a random number $\alpha\in[0;1]$
   \STATE Define interpolated patches $z^i_j=\alpha z^c_j+(1-\alpha)z_j^n$
   \FOR{$j=1$ {\bfseries to} $m$}
   \STATE $D_j(\theta)=r_\theta(z^n_j)-r_\theta(z^c_j)-\mu (||\nabla_z r_\theta(z^i_j)||_2-1)^2 $
   \ENDFOR
      \STATE $\theta\leftarrow Adam(\nabla_\theta \sum_{j=1}^mD_j(\theta))$
      \ENDFOR
\end{algorithmic}
\end{algorithm}

\subsection{Solving the variational problem}\label{sec:solve_var}
Image restoration is realized by solving the variational problem \eqref{eq:var}. To do so, we search for the minimizing image $x$  by performing $50$ iterations of Adam \cite{kingma2017adam}, with the momentum parameter set to the default values $\beta_1=0.9$ and $\beta_2=0.999$, and an exponential learning rate decay, with an initial learning rate of $0.1$ and a final learning rate of $0.01$ at the last iteration. We implement the method with the pytorch deep learning framework, so that the gradient of the global regularization functional $R(x)$ can be easily computed using automatic differentiation.

\section{Generalization to unseen noise level}\label{sec:generalization}
In this section, we study the robustness of the proposed regularization function to noise variations. The adversarial training of the regularization function, presented in the previous sections, requires to learn a different regularization function for every different noise level $\sigma$. We show how this limitation can be overcome.

We first analyze the behaviour of regularization functions trained on a single noise level $\sigma_{train}$ and then used to denoise an image with a different noise level $\sigma_{img}$. 
Second,  we propose to train the regularization functions with varying noise levels and demonstrate experimentally the superiority of this approach.

\subsection{Robustness to unseen noise level}
To study the ability of the local regularization function to generalize to noise levels unseen during training, we train 4 regularization functions on 4 different noise levels $\sigma_{train} \in \{0.05, 0.1, 0.2, 0.4\}$. We then evaluate the quality of the regularization of those networks on denoising tasks, for 5 different noise levels $\sigma_{img} \in \{0.05, 0.1, 0.2, 0.3, 0.4\}$. The 4 networks share the same architecture and the same training procedure as described in section \ref{sec:practical}. 

\begin{wrapfigure}[17]{r}{5.9cm}
\centering
\includegraphics[scale=0.33]{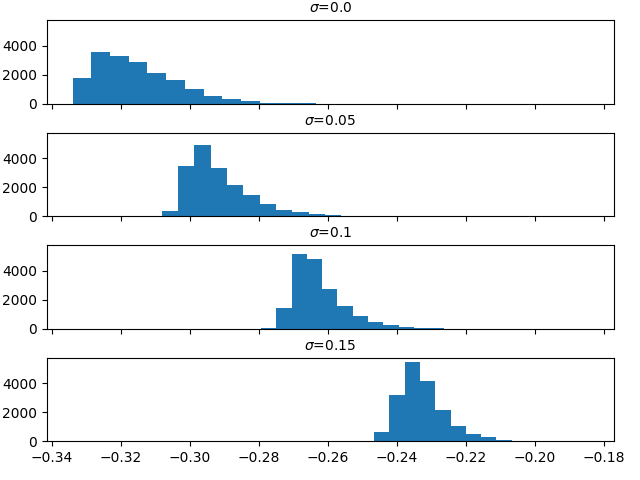}
\caption{Distribution of values $r_\theta(x)$ for a regularizer  trained on noise level $\sigma_{train}=0.1$. It generalizes well to patches $x$ with an intermediate ($0.05$) or extrapolated ($0.15$) noise levels.}
\label{fig:hist}
\end{wrapfigure}

While these regularizers have only been trained to distinguish between clean patches and noisy patches for a particular noise level, they generalize well to intermediate noise levels, in the sense that the regularizer value is an increasing function of the noise level of its input patch. Figure~\ref{fig:hist} illustrates this point for the noise level $\sigma_{train}=0.1$. The overlap between the distribution for noise level $\sigma=0$ (top) and $\sigma=0.1$ (bottom) is small, showing the ability of the regularizer network to distinguish clean and noisy patches. Furthermore, the distribution for noise level $\sigma=0.05$ is located in between the distributions $\sigma=0$ and $\sigma=0.1$, showing the ability of the network to generalize to intermediate noise levels but also to extrapolate to the noise level $0.15$.

\setlength{\columnsep}{17pt}


Next, we evaluate denoising quality by measuring the average PSNR on a validation set of $11$ images. 
To that end, we solve problem \eqref{eq:var} for $A=\mathrm{Id}$. We denoise images with 5 noise levels $\sigma_{img} \in\{0.05, 0.1, 0.2, 0.3, 0.4\}$, and we respectively set the regularization parameter $\lambda$ to $\{0.15, 0.35, 0.6, 0.8, 1\}$.
The results displayed on table \ref{tab:gen} demonstrate that the trained regularization functions generalize well to unseen noise level, as for all 5 levels of noise $\sigma_{img}$, the 4 regularization functions yield average PSNR values that are contained in an interval of size smaller than $1$ dB. Furthermore, regularization networks trained on small noise levels $\sigma_{train} \in \{0.05, 0.1\}$ generalize well to higher noise levels $\sigma_{train}$ as they perform even better than networks trained on the specific noise level.\\

We suggest that this is due to the fact that, when trained on a small noise level, the regularization function is forced to learn a tight boundary between the clean and the noisy distribution which favors denoising performance.
However, for the highest noise level $\sigma_{img}=0.4$, the regularization function trained on a small noise level $\sigma_{train}=0.05$ gives the worst results. 
As patches with very high noise levels are not seen during the training of the regularization function trained for $\sigma_{train}=0.05$, we suggest that the gradient penalty is not enforced to $1$ in this region of the patch space. Thus there is no guarantee that the gradient of the regularization function $\nabla r_{\theta}$ is indeed directed towards the space of possible solutions. This prevents the optimization algorithm from finding a relevant local minimum of \eqref{eq:var}.

%
\begin{table}[]
\centering
\caption{Average PSNR on AWGN denoising, in function of the image noise level $\sigma_{img}$, and the noise level the regularization network was trained on $\sigma_{train}$. For each image noise level, best result is displayed in bold, and second best result is underlined. Regularization networks trained on small noise level $\sigma_{train}$ generalizes well to higher noise levels $\sigma_{img}$. The regularization network trained on varying level of noise $\sigma_{train}\in[0.05;0.3]$ performs better on high noise levels $\sigma_{img}$.}
\begin{tabular}{c|c|c|c|c|c}
 \diagbox{$\sigma_{train}$}{$\sigma_{img}$} & 0.05 & 0.1 & 0.2 & 0.3 & 0.4 \\ 
\rowcolor{gray!20} 0.05 & \textbf{33.24} & \textbf{28.94} & \underline{24.21} & \underline{21.25} & 18.91  \\ 
 0.1 & \underline{33.20} & 28.82 & 24.17 & \underline{21.25} &  \underline{19.13}\\ 
\rowcolor{gray!20} 0.2 & 32.42 & 28.23 & 23.80 & 21.03 & {18.96}  \\ 
 0.4 & 33.01 & 28.23 & 23.84  & 21.03 & {18.96}  \\ 
 \rowcolor{gray!20} $[0.05; 0.3]$ & 32.92 & \underline{28.90} & \textbf{24.91} & \textbf{22.58} & \textbf{20.84} 
\end{tabular}
\vspace*{0.2cm}
\label{tab:gen}
\end{table}

\subsection{Robustness to noise variation during training}
We evaluate how robust the regularization function is to noise level variation during training. To do so, we train a regularization function on a distribution containing patches with noise level $\sigma_{train}$ uniformly sampled in the interval $[0.05, 0.30]$. We use the same network architecture and the same training procedure as in section \ref{sec:practical}. 

We evaluate the effectiveness of this regularization function by measuring the average PSNR when this function is used for denoising. We compare the performance with the prior trained on a single noise level in the last row of  table \ref{tab:gen}. Results show that the regularization function trained with a varying noise level has comparable performance with the regularization function trained on a single-noise level. Furthermore, for high noise level, the regularization function trained on a varying noise level significantly outperforms the regularization function trained on a single-noise level. This illustrates the fact that training the regularization function on varying noise level is actually beneficial. 

We suggest that exposing the regularization function to various noise levels during training combines two advantages.  It first learns a tight boundary around the clean patches distribution, as the networks trained on low noise levels. Second, the gradient-penalty is enforced even on highly noisy patches, as the networks trained on high noise levels.


\section{Experiments}\label{sec:experiment}

We evaluate the effectiveness of our learned regularization functional on two image restoration tasks, image denoising and image deblurring. 

\subsection{Denoising}
We evaluate our method on additive white Gaussian noise denoising, which corresponds to solving \eqref{eq:var} with $A=I$. We compare our method against two common patch-based denoising algorithms, BM3D \cite{BM3D} and EPLL \cite{EPLL}, on 3 noise levels $\sigma_{img} \in \{0.1, 0.2, 0.4\}$. We use our model trained on varying noise level $\sigma_{train} \in [0.05, 0.3]$, with the regularization parameter $\lambda$ respectively set to $0.15$, $0.35$ and $1$. For BM3D, we use the implementation of \cite{ipol-bm3d} with default parameters, and for EPLL we use the implementation of \cite{ipol-epll} with default parameters and a prior GMM model learned on RGB patches.

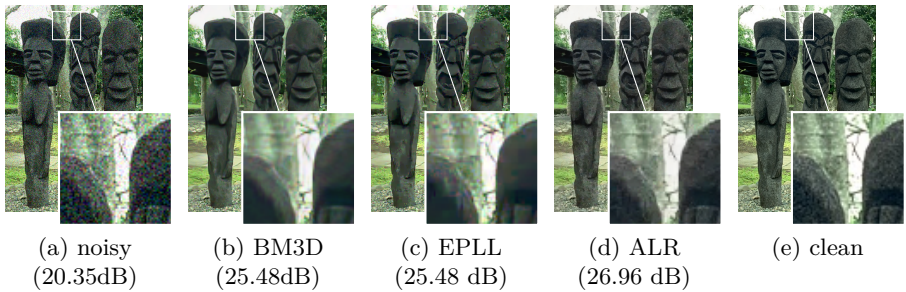
\begin{figure}
\centering
\begin{tabular}{ccccc}
		\begin{tikzpicture}[spy using outlines={rectangle, white,magnification=4, connect spies}]
		\node {\pgfimage[interpolate=true,width=.15\linewidth]{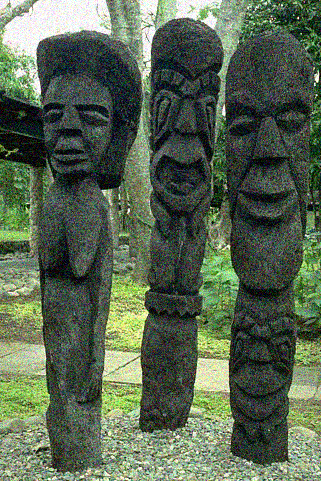}};
			\coordinate (spypoint) at (-0.1, 1.1);
		\coordinate (spyviewer) at (0.55,-0.8);
		\spy[width=1.5cm,height=1.5cm]on (spypoint) in node [fill=white] at (spyviewer);
		\end{tikzpicture}
		&
				\begin{tikzpicture}[spy using outlines={rectangle, white,magnification=4, connect spies}]
		\node {\pgfimage[interpolate=true,width=.15\linewidth]{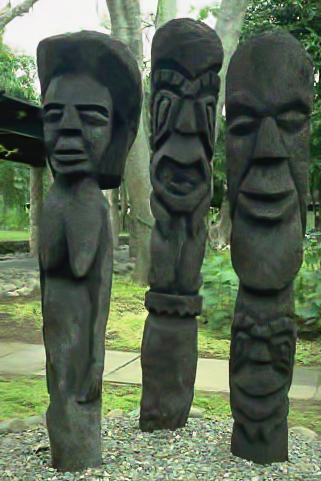}};
		\coordinate (spypoint) at (-0.1, 1.1);
		\coordinate (spyviewer) at (0.55,-0.8);
		\spy[width=1.5cm,height=1.5cm] on (spypoint) in node [fill=white] at (spyviewer);
		\end{tikzpicture}
		&
			\begin{tikzpicture}[spy using outlines={rectangle, white,magnification=4, connect spies}]
		\node {\pgfimage[interpolate=true,width=.15\linewidth]{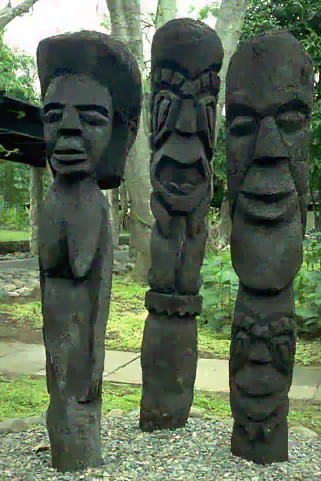}};
			\coordinate (spypoint) at (-0.1, 1.1);
		\coordinate (spyviewer) at (0.55,-0.8);
		\spy[width=1.5cm,height=1.5cm]on (spypoint) in node [fill=white] at (spyviewer);
		\end{tikzpicture}
		&
				\begin{tikzpicture}[spy using outlines={rectangle, white,magnification=4, connect spies}]
		\node {\pgfimage[interpolate=true,width=.15\linewidth]{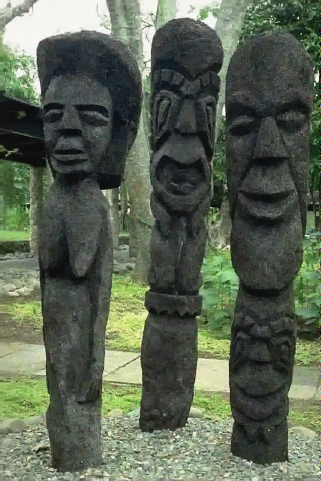}};
			\coordinate (spypoint) at (-0.1, 1.1);
		\coordinate (spyviewer) at (0.55,-0.8);
		\spy[width=1.5cm,height=1.5cm]on (spypoint) in node [fill=white] at (spyviewer);
		\end{tikzpicture}
		&
				\begin{tikzpicture}[spy using outlines={rectangle, white,magnification=4, connect spies}]
		\node {\pgfimage[interpolate=true,width=.15\linewidth]{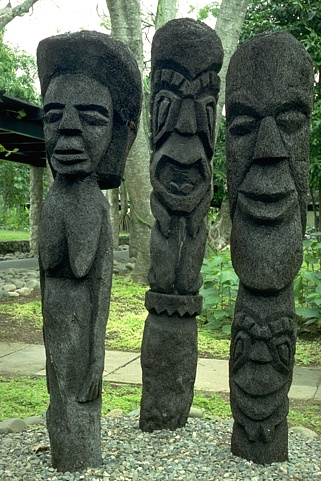}};
		\coordinate (spypoint) at (-0.1, 1.1);
		\coordinate (spyviewer) at (0.55,-0.8);
		\spy[width=1.5cm,height=1.5cm]on (spypoint) in node [fill=white] at (spyviewer);
		\end{tikzpicture}
		\\
		(a) noisy  &(b) BM3D & (c) EPLL&(d) ALR &(e) clean\\
		(20.35dB)&(25.48dB)&(25.48 dB)&(26.96 dB)
	\end{tabular}
\caption{Visual comparison of patch-based denoising methods for $\sigma=0.1$.}
\label{fig:awgn}
\end{figure}

The average PSNR and LPIPS \cite{zhang2018unreasonable} on the BSD68 dataset for the 3 methods are presented in Table \ref{tab:awgn}, and examples of denoised images are shown on Figure \ref{fig:awgn}. For the 3 noise levels, the adversarial local regularization denoising outperform EPLL and BM3D in terms of PSNR, while having comparable perceptual quality. This illustrates the ability of convolutional neural networks to be used as local regularizers when trained the right way.

\begin{table}[]
    \caption{Comparisons in terms of PSNR (left) and LPIPS (right) of the patch-based denoising algorithms ALR,  EPLL and BM3D, for white Gaussian noise. Results are averaged on the 68 images of the BSD68 dataset. \label{tab:awgn}}
      \centering
        \begin{tabular}{c||c|c|c||c|c|c}
        \multicolumn{1}{c}{}&\multicolumn{3}{c||}{PSNR}&\multicolumn{3}{c}{LPIPS}\\
          $\sigma$ &  ALR & EPLL & BM3D&ALR & EPLL & BM3D   \\ 
         \rowcolor{gray!20}   0.1 & {\bf 28.85} & \underline{28.77} & 28.26& \underline{ 0.29} & {\bf 0.28} &  0.30 \\ 
            0.2 & \underline{24.88} & {\bf 24.92} & 24.69&  0.44 & {\bf 0.42} & \underline{0.43}\\ 
         \rowcolor{gray!20}   0.4 & {\bf 21.58 }& 19.75 &\underline{20.25} & {\bf 0.57} & 0.61 & \underline{0.58}\\ 
        \end{tabular}
        \vspace*{0.2cm}
\end{table}

\subsection{Deblurring}
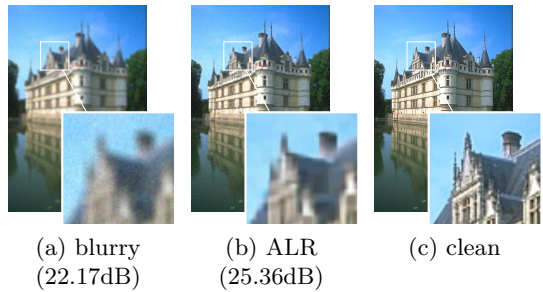
\begin{figure}[h!]
\centering
\begin{tabular}{ccc}
		\begin{tikzpicture}[spy using outlines={rectangle, white,magnification=4, connect spies}]
		\node {\pgfimage[interpolate=true,width=.15\linewidth]{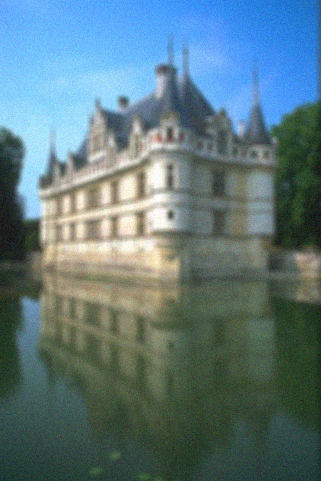}};
			\coordinate (spypoint) at (-0.3, 0.7);
		\coordinate (spyviewer) at (0.55,-0.8);
		\spy[width=1.5cm,height=1.5cm]on (spypoint) in node [fill=white] at (spyviewer);
		\end{tikzpicture}
		&
				\begin{tikzpicture}[spy using outlines={rectangle, white,magnification=4, connect spies}]
		\node {\pgfimage[interpolate=true,width=.15\linewidth]{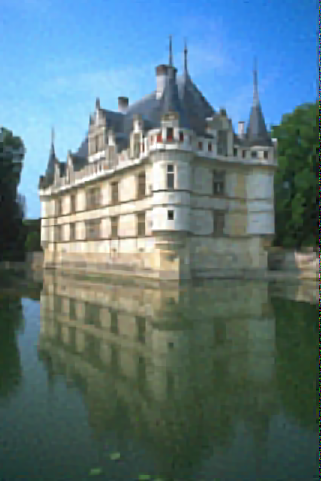}};
		\coordinate (spypoint) at (-0.3, 0.7);
		\coordinate (spyviewer) at (0.55,-0.8);
		\spy[width=1.5cm,height=1.5cm] on (spypoint) in node [fill=white] at (spyviewer);
		\end{tikzpicture}
		&
			\begin{tikzpicture}[spy using outlines={rectangle, white,magnification=4, connect spies}]
		\node {\pgfimage[interpolate=true,width=.15\linewidth]{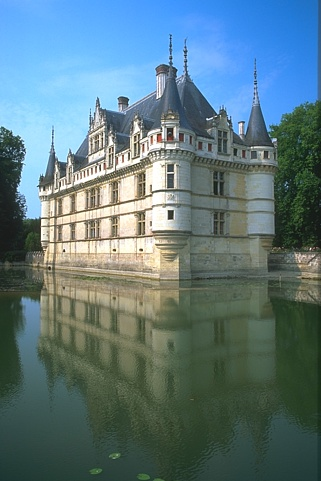}};
			\coordinate (spypoint) at (-0.3, 0.7);
		\coordinate (spyviewer) at (0.55,-0.8);
		\spy[width=1.5cm,height=1.5cm]on (spypoint) in node [fill=white] at (spyviewer);
		\end{tikzpicture}
		\\
		(a) blurry &(b) ALR & (c) clean \\
		(22.17dB)&(25.36dB)& 
	\end{tabular}
\caption{Illustration of deblurring using a $7 \times 7$ Gaussian  kernel with standard deviation  $\sigma_{k} = 3$.}
\label{fig:deb}
\end{figure}

To illustrate the adaptability of our local regularization function, we consider image deblurring. 
This corresponds to solving \eqref{eq:var} with a linear degradation operator $A$ taken as a convolution operation with a blur kernel $k$, that is $y = k * x + \varepsilon$. Figure \ref{fig:deb} shows an example of image deblurring using our learned local regularization function. The image is blurred with a $7 \times 7$ Gaussian  kernel with standard deviation  $\sigma_{k} = 3$, and an additive  white Gaussian noise of standard deviation $\sigma=0.03$. 

\section{Conclusion and Perspectives}

We propose a new strategy to solve inverse problem in imaging using a convolutional neural network as a local regularization function. The local regularization network is trained to discriminate between clean and noisy patches, and the global regularization function is defined as the average value of the local function over the set of all image patches. Working with a local regularization function offers several advantages : it works with any image size, it requires less training data and has less parameters than a full size model. Furthermore, the fully convolutional architecture of the network makes it computationally efficient to compute the global regularization function and its gradient.

Experimental results on image denoising show that our method outperforms popular patch-based denoising algorithm such as EPLL and BM3D, illustrating the potential of convolutional networks to acts as regularization function for inverse imaging problems. 

We believe that improving the training criterion of the regularization function could improves the performance of the regularization. Indeed, the training criterion of our local regularization network corresponds to the 1-Wasserstein distance. The regularizer thus  grows linearly with the distance to the clean data manifold, whereas the data-fidelity term is quadratic. We suggest that these unbalanced terms make  the variational problem difficult to solve, especially for high noise levels. We believe that learning a regularization term based on the 2-Wasserstein distance could help to overcome this limitation, as the learned regularization function would then grow with the square of the distance to the clean manifold.



\vspace*{-0.2cm}

\paragraph{Acknowledgements.}
This study has been carried out with financial support from the French Research Agency through the PostProdLEAP project (ANR-19-CE23-0027-01).\vspace*{-0.3cm}

\bibliographystyle{splncs04}
\bibliography{refs}

\begin{thebibliography}{10}
\providecommand{\url}[1]{\texttt{#1}}
\providecommand{\urlprefix}{URL }
\providecommand{\doi}[1]{https://doi.org/#1}

\bibitem{WGAN}
Arjovsky, M., Chintala, S., Bottou, L.: Wasserstein generative adversarial
  networks. In: International conference on machine learning. pp. 214--223.
  PMLR (2017)

\bibitem{Bigdeli2017}
Bigdeli, S.A., Jin, M., Favaro, P., Zwicker, M.: {Deep Mean-Shift Priors for
  Image Restoration}. In: Adv. in Neural Information Proces. Systems 30. pp.
  763--772 (2017)

\bibitem{bora2017compressed}
Bora, A., Jalal, A., Price, E., Dimakis, A.G.: Compressed sensing using
  generative models. In: Int. Conference on Machine Learning. vol.~2, pp.
  537--546 (2017)

\bibitem{Coifman1995}
Coifman, R.R., Donoho, D.L.: {Translation-Invariant De-Noising}. In: Wavelets
  and Statistics (Lect. Notes in Statistics, vol 103), chap.~5, pp. 125--150.
  Springer (1995)

\bibitem{BM3D}
Dabov, K., Foi, A., Katkovnik, V., Egiazarian, K.: Image denoising by sparse
  3-d transform-domain collaborative filtering. IEEE Trans. on Image Processing
   \textbf{16},  2080--95 (2007)

\bibitem{Donoho1994}
Donoho, D.L., Johnstone, J.M.: {Ideal spatial adaptation by wavelet shrinkage}.
  Biometrika  \textbf{81}(3),  425--455 (1994)

\bibitem{Gonzalez2019}
Gonz{\'{a}}lez, M., Almansa, A., Delbracio, M., Mus{\'{e}}, P., Tan, P.:
  {Solving Inverse Problems by Joint Posterior Maximization with a VAE Prior}
  (2019)

\bibitem{GAN}
Goodfellow, I.J., Pouget-Abadie, J., Mirza, M., Xu, B., Warde-Farley, D.,
  Ozair, S., Courville, A., Bengio, Y.: Generative adversarial networks. arXiv
  preprint arXiv:1406.2661  (2014)

\bibitem{WGANGP}
Gulrajani, I., Ahmed, F., Arjovsky, M., Dumoulin, V., Courville, A.: Improved
  training of wasserstein gans. In: Adv. in Neural Information Processing
  Systems. vol.~30, pp. 5767--5777 (2017)

\bibitem{HDMI}
Houdard, A., Bouveyron, C., Delon, J.: High-dimensional mixture models for
  unsupervised image denoising ({HDMI}). SIAM J. Imag. Sc.  \textbf{11}(4),
  2815--2846 (2018)

\bibitem{ipol-epll}
Hurault, S., Ehret, T., Arias, P.: {EPLL: An Image Denoising Method Using a
  Gaussian Mixture Model Learned on a Large Set of Patches}. {Image Processing
  On Line}  \textbf{8},  465--489 (2018)

\bibitem{kamilov2017plug}
Kamilov, U.S., Mansour, H., Wohlberg, B.: A plug-and-play priors approach for
  solving nonlinear imaging inverse problems. IEEE Signal Processing Letters
  \textbf{24}(12),  1872--1876 (2017)

\bibitem{kingma2017adam}
Kingma, D.P., Ba, J.: Adam: A method for stochastic optimization. In:
  International Conference on Learning Representations (2015)

\bibitem{ipol-bm3d}
Lebrun, M.: {An Analysis and Implementation of the BM3D Image Denoising
  Method}. {Image Processing On Line}  \textbf{2},  175--213 (2012)

\bibitem{lunz2019}
Lunz, S., \"{O}ktem, O., Sch\"{o}nlieb, C.B.: Adversarial regularizers in
  inverse problems. In: Adv. in Neural Information Processing Systems. vol.~31,
  pp. 8507--8516 (2018)

\bibitem{meinhardt2017learning}
Meinhardt, T., Moller, M., Hazirbas, C., Cremers, D.: Learning proximal
  operators: Using denoising networks for regularizing inverse imaging
  problems. In: International Conference on Computer Vision. pp. 1781--1790
  (2017)

\bibitem{nair2010rectified}
Nair, V., Hinton, G.E.: Rectified linear units improve restricted boltzmann
  machines. In: International Conference on Machine Learning (2010)

\bibitem{Reehorst2018a}
Reehorst, E.T., Schniter, P.: {Regularization by Denoising: Clarifications and
  New Interpretations}. IEEE Trans. on Computational Imaging  \textbf{5}(1),
  52--67 (mar 2019)

\bibitem{Romano2016red}
Romano, Y., Elad, M., Milanfar, P.: {The Little Engine That Could:
  Regularization by Denoising (RED)}. SIAM J. on Imaging Sciences
  \textbf{10}(4),  1804--1844 (2017)

\bibitem{rudin1992nonlinear}
Rudin, L.I., Osher, S., Fatemi, E.: Nonlinear total variation based noise
  removal algorithms. Physica D: nonlinear phenomena  \textbf{60}(1-4),
  259--268 (1992)

\bibitem{ryu2019plug}
Ryu, E.K., Liu, J., Wang, S., Chen, X., Wang, Z., Yin, W.: Plug-and-play
  methods provably converge with properly trained denoisers. In: International
  Conference on Machine Learning. pp. 5546--5557 (2019)

\bibitem{santambrogio2015ot}
Santambrogio, F.: Optimal transport for applied mathematicians. Progress in
  Nonlinear Differential Equations and their applications  \textbf{87} (2015)

\bibitem{Teodoro2018scene}
Teodoro, A.M., Bioucas-Dias, J.M., Figueiredo, M.A.T.: {Scene-Adapted
  Plug-and-Play Algorithm with Guaranteed Convergence: Applications to Data
  Fusion in Imaging} (2018)

\bibitem{venkatakrishnan2013plug}
Venkatakrishnan, S.V., Bouman, C.A., Wohlberg, B.: Plug-and-play priors for
  model based reconstruction. In: IEEE Global Conference on Signal and
  Information Processing. pp. 945--948 (2013)

\bibitem{weyand2020google}
Weyand, T., Araujo, A., Cao, B., Sim, J.: Google landmarks dataset v2 -- a
  large-scale benchmark for instance-level recognition and retrieval (2020)

\bibitem{Yu2011}
Yu, G., Sapiro, G.: {DCT image denoising: a simple and effective image
  denoising algorithm}. Image Processing On Line  \textbf{1} (2011)

\bibitem{Zhang2017}
Zhang, K., Zuo, W., Gu, S., Zhang, L.: {Learning Deep CNN Denoiser Prior for
  Image Restoration}. In: IEEE Conf. Comput. Vis. Pat. Recog. pp. 2808--2817
  (2017)

\bibitem{zhang2018unreasonable}
Zhang, R., Isola, P., Efros, A.A., Shechtman, E., Wang, O.: The unreasonable
  effectiveness of deep features as a perceptual metric (2018)

\bibitem{EPLL}
Zoran, D., Weiss, Y.: From learning models of natural image patches to whole
  image restoration. In: Int. Conference on Computer Vision. pp. 479--486
  (2011)

\end{thebibliography}

\end{document}